\documentclass[oldversion]{aa} 
\usepackage{graphicx}
\usepackage{txfonts}
\usepackage{natbib}
\bibpunct{(}{)}{;}{a}{}{,}

\pdfoutput=0

\begin{document}
 
\title{Discovery of a nearby young brown dwarf binary candidate}

\author{A.~Reiners
  \inst{1}\fnmsep\thanks{Emmy Noether Fellow},
 A.~Seifahrt\inst{2}
 \and
 S.~Dreizler\inst{1}
}


\institute{
  Universit\"at G\"ottingen, Institut f\"ur Astrophysik, Friedrich-Hund-Platz 1, D-37077 G\"ottingen, Germany\\
  \email{Ansgar.Reiners@phys.uni-goettingen.de}
  \and
  Department of Physics, University of California, One Shields Avenue, Davis, CA 95616, USA
  }

\date{Accepted Mar 05, 2010}


\abstract {In near-infrared NaCo observations of the young brown dwarf
  2MASS~J0041353-562112, we discovered a companion a little less than
  a magnitude fainter than the primary.  The binary candidate has a
  separation of 143\,mas, the spectral types are M6.5 and M9.0 for the
  two components.  Colors and flux ratios are consistent with the
  components being located at the same distance minimizing the
  probability of the secondary being a background object. The brown
  dwarf is known to show Li absorption constraining the age to less
  than $\sim200$\,Myr, and it was suspected to show ongoing accretion,
  indicating an age as low as $\sim 10$\,Myr.  We estimate distance
  and orbital parameters of the binary as a function of age. For an
  age of 10\,Myr, the distance to the system is 50\,pc, the orbital
  period is 126\,yr, and the masses of the components are $\sim30$ and
  $\sim15$\,M$_{\rm{Jup}}$.  The binary brown dwarf fills a so far
  unoccupied region in the parameters mass and age; it is a valuable
  new benchmark object for brown dwarf atmospheric and evolutionary
  models.  }


\maketitle
%

\section{Introduction}
\label{sect:Introduction}

Very low mass binaries are of particular interest for a number of
reasons. First, the binary fraction at very low masses as well as
their orbital properties carry important information about the way
binaries form \citep[e.g.][]{Close03, Burgasser07}. Second, all
components of a multiple system share the same evolutionary history so
that a comparison between binary components is free of a number of
degeneracies. Finally, binaries offer a model-independent way to
determine the mass through measurement of their orbital period. This
third point is most important for our understanding of the evolution
of low-mass stars and brown dwarfs in particular at young ages where
models still have relatively large uncertainties.

The number of very low mass binaries has grown rapidly over the last
years. An updated list of binaries with total masses below
0.2\,M$_\odot$ can be found at \texttt{www.vlmbinaries.org}, this list
carries 99 entries as of Jan~2010. Of particular importance are the
brown dwarf binaries with independent age constraints because they
deliver empirical constraints on brown dwarf evolution models.
Furthermore, in order to determine the mass from orbital motion on
reasonable timescales, the orbital period should be short enough, and
the binaries should not be too far away so that spectroscopic
investigation is possible. Usually, young binaries are members of star
forming regions that are located at a distance of 100\,pc or more,
which makes a detailed investigation of low-mass members very
difficult. Therefore, young, nearby, low-mass systems are of very
great value for our understanding of low-mass star and brown dwarf
evolution.

In this paper, we present the discovery of a new very low mass binary,
2MASS~0041353-562112 (hereafter 2M0041), that is nearby and probably
very young. The age of 2M0041 is constrained to be lower than
$\sim200$\,Myr by the detection of Li in an optical spectrum
\citep{Reiners09b}. \citet{Reiners09} presents evidence for accretion
deduced from the intensity and shape of emission lines, in particular
H$\alpha$. Ongoing accretion would suggest that 2M0041 may even be as
young as $~10$\,Myr. Space motion of 2M0041 is consistent with it
being a member of the $\sim20$\,Myr old Tuc-Hor association or the
$\sim12$\,Myr old $\beta$~Pic association.  Unfortunately, the
distance to 2M0041 is not yet known because no parallax measurement is
available so far. The distance from spectrophotometry would be 17\,pc
if the object was an old, single field star, but the real distance is
of course larger because the young object is more luminous. As long as
no parallax is measured for 2M0041, age, mass, and distance are free
parameters that can be constrained by measuring the orbital period of
the system.

\section{Data and Analysis}
\label{sect:Data}

\begin{figure}
  \center
  \resizebox{.9\hsize}{!}{\includegraphics[]{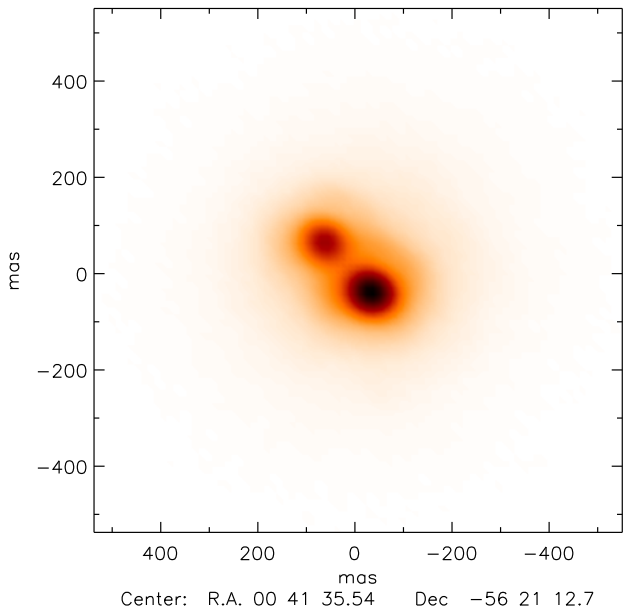}}
  \caption{NACO K-band image of 2M0041, North is up and East is left.
    Axis labels denote coordinates in mas relative to the center of
    the image.\label{fig:Image}}
\end{figure}

Data were obtained with NaCo, the Nasmyth Adaptive Optics System
(NAOS) and Near-Infrared Imager and Spectrograph (CONICA) at ESO's
Very Large Telescope \citep{Lenzen03, Rousset03}. Four images were
obtained in service mode on August 14, 2009: One image was taken in
$J$, one in $H$, and two images were obtained in $K_s$. We obtained
ten individual frames, each of them being the average of 3
observations with an individual exposure time of 30\,s each. Thus, the
total exposure time is 15\,min per image. Individual images were
jittered using a 5\arcsec\ jitter box, this allowed efficient
reduction of the sky background. The adaptive optics system was used
with the N90C10 dichroic using 90\% of the light for AO and only 10\%
for the science camera. We chose this option because no bright AO
source was available nearby. All observations were carried out at low
airmass ($<1.3$) and good seeing conditions ($<0.8\arcsec$). We used
the S13 camera with a field of view of $14\arcsec \times 14\arcsec$
and a pixel scale of 13.2\,mas/pix. For our analysis, we use the
standard pipeline products that are provided for service mode
observations. Data reduction includes dark subtraction, flat fielding,
and sky subtraction using the jittered images. A close-up of one of
the two $K_s$-band images is shown in Fig.\,\ref{fig:Image}. 2M0041 is
clearly resolved as a binary with two components that are somewhat
different in brightness; the component to the SW appears brighter than
the NE component.

To measure the separation and the flux ratio of the two components, we
estimate the PSF from the science images in an iterative procedure.
We start with an image of a standard star. We use the so-called
zero-point images that are routinely taken for NaCo service
observations. With this first guess of the PSF, we search for stars in
our science images and determine their position and flux using the
public domain IDL package \emph{StarFinder} \citep{Diolaiti00}. Next,
we extract the PSF from the science image given the information on
flux and position using the routine \texttt{psf\_extract} from the
\emph{StarFinder} package. We use this PSF in the next step to
iteratively search for the two components again and redetermine the
shape of the PSF.

The deduction of the PSF from the science image has the advantage that
we do not have to rely on a PSF that was taken at a different time,
direction, and object brightness. The shape of the PSF sensitively
depends on the adaptive optics performance which depends on the
brightness of the reference target. On the other hand, the separation
of the two components that we wish to distinguish is not much larger
than the width of the PSF, particularly in the $J-$band. During
determination of the PSF, this can lead to the problem that the
algorithm constructs a PSF that consists of both components. To avoid
this problem, we assume an axisymmetric shape of the PSF. After each
PSF determination, we construct a rotationally averaged version of the
PSF that we use for the next iteration of \emph{StarFinder}.

We successfully identified the two components in all four images using
the procedure described above. As results, we obtain the positions and
individual flux of the two components, and the PSF of each
image.

\begin{table}
  \center
  \caption{\label{tab:measurements} Separation and flux measurements in the four images.}
  \begin{tabular}{cccccc}
    \hline
    \hline
    \noalign{\smallskip}
    \# & Band & Separation & PA & Flux ratio & PSF width\\
    && [mas] &&& [mas]\\
    \noalign{\smallskip}
    \hline
    \noalign{\smallskip}
    1 & $J$   & 142.3 & $44.90^\circ$ & $2.19 \pm 0.11$ & 101.1 \\
    2 & $H$   & 142.4 & $43.34^\circ$ & $2.06 \pm 0.06$ &  72.3 \\
    3 & $K_s$ & 143.0 & $43.46^\circ$ & $1.88 \pm 0.05$ &  48.2 \\
    4 & $K_s$ & 143.4 & $43.25^\circ$ & $1.88 \pm 0.05$ &  52.9 \\
    \noalign{\smallskip}
    \hline
  \end{tabular}
\end{table}

We show the separation of the two components, their flux ratio, and
the width of the PSF from our four images in
Table\,\ref{tab:measurements}. The results for the separation are
consistent with each other. The flux ratios from the two $K_s$-band
images agree very well, and the flux ratio seems to be a function of
color, which is consistent with two components of different
temperature. The width of the PSF also varies with wavelength, this is
expected because the Strehl ratio is lower at shorter wavelengths
(Strehl ratios are 0.50 in $K_s$, 0.33 in $H$, and 0.18 in $J$). Note
that in the $K_s$-band the width of the PSF is on the order of the
diffraction limit. Going to even longer wavelengths would not lead to
better image quality because the diffraction limit grows with longer
wavelengths (for example, it is $\sim100$\,mas at $L'$). In the
$J$-band, the two components are not very well separated. We tried to
use a PSF consisting of different elliptical components, but did not
succeed to produce results better than our procedure described
above. The $J$-band position angle differs from the $H$- and
$K_s$-band results by about 1.5 degree, which can be attributed to the
slightly irregular shape of the $J$-band PSF. In particular,
irregularities in the wings of the $J$-band PSF may lead to a small
shift of the suspected stars' positions (here, $\la$0.2 pix or
$\sim$2.5\% of the $J$-band FWHM). This probably leads to a slight
dependence of the photocenter on the stars intensity in the $J$-band.

To quantify the uncertainty of our measurements, we performed
Monte-Carlo simulations in all three bands.  In each band, we
constructed a series of 1000 images of a binary consisting of two
objects with different flux ratios, separations, and position
angles. We uniformly varied these values around those found in the
real data, and we used asymmetric PSFs consisting of two elliptical
Gaussian components that scatter around a description optimized to fit
the rotationally symmetric PSF from the procedure described
above. Noise was added to the artificial data in order to match the
quality of the real images. In the $H$- and $K$-bands, the fitting
process turned out to be very stable. The clear separation of the two
components due to the narrow PSF allows a robust fitting process even
in the presence of an asymmetric PSF. We derive the uncertainties in
the flux ratio from the scatter around the mean of our
simulations. The 2$\sigma$-scatter of the flux ratio is 3\% in $H$ and
2\% in $K_s$. The $J$-band uncertainties are larger. The main reason
is not only the larger FWHM of the PSF, but also the asymmetry in the
PSF mentioned above. Thus, the fitting process in the $J$-band image
sensitively depends on assumptions on the shape of the PSF. Because
deconvolving that shape is less reliable in the $J$-band image, we
decided not to simply adopt the $J$-band uncertainties from our
Monte-Carlo approach, but conservatively estimate them to be roughly a
factor of two larger because of PSF-dependent systematics. We adopt a
$J$-band flux ratio uncertainty of 5\%, which well covers the scatter
we found during our attempts using different PSF assumptions.

\section{Photometry of the components}
\label{sect:Photometry}

Photometric measurements of 2M0041 in $J$, $H$, and $K_s$ are
available from the Two Micron All Sky Survey \citep[2MASS,][]{2MASS}.
The 2MASS magnitude of 2M0041 reflects the combined flux from both
components. With the NaCo observations, we can now separate the
combined flux into the two components using the 2MASS magnitudes and
the flux ratios $f_1/f_2$ in the three filters; for the $J$-band we
can write

\begin{equation}
  J_2 = J + 2.5 \log{\left[1 + \left(\frac{f_1}{f_2}\right)_J\right]}.
\end{equation}

The results for individual $J$, $H$, and $K_s$ magnitudes are
summarized in Table\,\ref{tab:photometry}. Under the assumption that
the two components are indeed forming a binary (i.e., both are located
at the same distance $d$), we can determine the individual spectral
types from the magnitude differences and from the information about
the combined spectral type from integrated light. Relations between
absolute $J$-magnitude and Spectral Types were reported by
\citet{Dahn02} and \citet{Cruz03}. The combined spectral type of
2M0041 is M7.5 with an uncertainty of $\pm0.5$
\citep{PhanBao06}. Using the linear relation from Dahn et al. we find
that the Spectral Type difference between the two individual
components is 2.5 spectral classes. The dispersion of objects around
the relation from Dahn et al. is 0.25~mag, which translates into an
individual uncertainty of 0.7 spectral classes for individual objects,
and an uncertainty of 1.0 spectral classes for the difference between
two objects. This dispersion is much larger than the uncertainty from
the photometric differences of 2M0041~A and B so that we can neglect
the latter.

\begin{table}
  \center
  \caption{\label{tab:photometry} Measured parameters of the system}
  \begin{tabular}{cccc}
    \hline
    \hline
    \noalign{\smallskip}
    & 2MASS & \multicolumn{2}{c}{Individual photometry} \\
    Parameter & A+B & A & B\\
    \noalign{\smallskip}
    \hline
    \noalign{\smallskip}
    $J$   & $11.96 \pm 0.02$ & $12.37 \pm 0.03$ & $13.22 \pm 0.04$ \\
    $H$   & $11.32 \pm 0.02$ & $11.75 \pm 0.02$ & $12.53 \pm 0.03$ \\
    $K_s$ & $10.86 \pm 0.03$ & $11.32 \pm 0.03$ & $12.01 \pm 0.03$ \\
    \noalign{\smallskip}
    $J-K_s$ & $1.10 \pm 0.04$ & $1.05 \pm 0.04$ & $1.21 \pm 0.05$ \\
    \noalign{\smallskip}
    $\Delta J$ && \multicolumn{2}{c}{$0.85 \pm  0.04$}\\
    $\Delta H$ && \multicolumn{2}{c}{$0.78 \pm  0.02$}\\
    $\Delta K_s$ && \multicolumn{2}{c}{$0.69 \pm  0.02$}\\
    \noalign{\smallskip}
    SpT   && M$6.5 \pm 1$ & M$9.0 \pm 1$\\
    \noalign{\smallskip}
    Position Angle && \multicolumn{2}{c}{$43.6\degr \pm 0.6\degr$}\\
    Separation [mas] && \multicolumn{2}{c}{$142.8 \pm 0.5$}\\
    \noalign{\smallskip}
    \hline
  \end{tabular}
\end{table}

The linear relation between Spectral Type and absolute magnitude also
allows us to use the combined spectral type (M7.5) to anchor the
spectral type range of 2M0041~A and B. Using flux ratio and individual
magnitudes, we can construct an artifical average magnitude of the
combined system 2M0041AB, ($\tilde{M}_{J\rm{, AB}} = 12.63$,
$\tilde{M}_{K\rm{, AB}} = 11.55$), which in this system must coincide
with Spectral Type M$7.5 \pm 0.5$. From that, we derive individual
Spectral Types for 2M0041~A and B of M$6.5 \pm 1.0$ and M$9.0 \pm
1.0$, respectively.

An independent check of our assumption that both components are indeed
located at the same distance comes from the colors of 2M0041~A and
B. $J-K$ colors of late-M objects are given, e.g., in
\citet{Hawley02}, \citet{Dahn02}, and \citet{West08}. The colors of
2M0041~A indicate a spectral type in the range M5--M9 while 2M0041~B
falls in the range M8.5--early L. Thus, the colors and flux ratios of
2M0041~A and B support the assumption that both components belong to a
binary system: Given a dispersion of 0.2\,mag in the $J-K$ relation
and much smaller uncertainties in our measurements of flux ratios and
colors, the difference in distance modulus between the two components
is unlikely to be larger than 6\,pc.

Table\,\ref{tab:photometry} also contains the mean position angle and
separation. For the position angle, the individual measurements are
weighted according to the inverse PSF widths of the images; the result
is $43.6\degr \pm 0.6\degr$. The mean separation calculated from the
four exposures is 142.8\,mas, we use the scatter of 0.5\,mas as a
conservative estimate for the uncertainty of this value. 

The proper motion of 2M0041 is $\sim140$\,mas\,yr$^{-1}$
\citep{PhanBao06}, i.e., the primary travels roughly by one full
observed separation per year in SE direction. The secondary is still
bright enough to be visible in archive nIR measurements if the
separation to the primary was large enough. We can use nIR or red
observations of the region to look whether any signatures of a second
object can be found. Assuming that the secondary would show negligible
proper motion, the objects should be separated by $\sim 1.5$\arcsec\
on the 2~MASS images taken in 1999, which would be difficult to detect
given the $\sim 3$\arcsec\ FWHM of the 2MASS PSF. On the other hand,
ESO.R-MAMA plates taken in 1988 should show two objects separated by
about one FWHM, which also is 2.5--3\arcsec. We found no second object
at the position of 2M0041 in the 2MASS images from 1999 and in the
ESO.R-MAMA images from 1988, and we see no signs of an elongated PSF
which could be indicative of a barely resolved second object close to
the primary ($\sim1$\arcsec). We conclude that it is very likely that
the second object in our images is physically bound to the
primary. Confirmation of binary status, however, can only be
accomplished by verifying common proper motion in an exposure taken at
a second epoch.

\section{System parameters}
\label{sect:System}

From the new photometry, we can determine the parameters of the two
components, and we can estimate mass and orbital period for a given
age. The distance to 2M0041 can be estimated from the difference
between absolute and apparent magnitude. For field objects, the
absolute magnitude as a function of spectral type was given by
\citet{Dahn02} and \citet{Cruz03}. We can use that estimate for the
lower limit of the distance, which would apply if the objects were
field stars. To calculate the distance that would apply if the object
was very young, we can then scale the absolute magnitude by the radius
difference between young and old objects assuming that the temperature
is not significantly changing with age for a given spectral type. We
use the radius-age dependence from \citet{Baraffe98, Baraffe02}.

All parameters together with uncertainties are given in
Table\,\ref{tab:parameters}. Uncertainties are due to the spectral
type uncertainty, which has an important effect on the absolute
magnitude, and to this we add the uncertainty in $J$ in quadrature.
Under the assumption that 2M0041~A is a field star, the distance to
the system is 24\,pc. As expected, the new distance is larger than the
17\,pc calculated earlier \citep{Faherty09}, because in that
calculation, all the flux measured in $J$ was assumed to come from a
single (and cooler) object.  We can now use the radius-age relation to
estimate the distance for different ages of 2M0041~A
\citep[see][]{Reiners09}. The distances to 2M0041 for ages of 5 and
10\,Myr are 71 and 50\,pc, respectively. To estimate the semi-major
axis, we apply a correction factor of 1.26 to account for projection
effects \citep{Fischer92}. The resulting separations between 2M0041~A
and B come out as 8.9 and 12.8\,AU for a 10\,Myr and a 5\,Myr binary,
respectively.

From the evolutionary tracks of Baraffe et al., we can also estimate
the mass of the two components at a given age. With the masses and the
separation known, we can then estimate the orbital period of the
binary assuming that the distance we observe is the true
semimajor-axis of the system.

\begin{table}
  \center
  \caption{\label{tab:parameters} System parameters for different ages. The main sequence (MS) case is excluded by the Li detection and given only for reference.}
  \begin{tabular}{ccc|c}
    \hline
    \hline
    \noalign{\smallskip}
    Parameter & \multicolumn{3}{c}{Value}\\
    Age [Myr] & 5 & 10 & MS \\
    \noalign{\smallskip}
    \hline
    \noalign{\smallskip}
    $d$ [pc] & $71^{+29}_{-15}$ & $50^{+21}_{-10}$ & $24^{+10}_{-5}$\\[1mm]
    Separation [AU]  & $12.8^{+5.3}_{-2.7}$ & $8.9^{+3.7}_{-1.9}$ & $4.3^{+1.8}_{-0.9}$\\[1mm]
     \noalign{\smallskip}
    Mass (A) [M$_{\rm Jup}$] & $\sim 25$ & $\sim 30$ & $\sim 95$\\
    Mass (B) [M$_{\rm Jup}$] & $\sim 15$ & $\sim 15$ & $\sim 80$\\[1mm]
    \noalign{\smallskip}
    Orbital Period [yr]  & $228^{+155}_{-68}$ & $126^{+86}_{-37}$ & $22^{+15}_{-7}$\\
    \noalign{\smallskip}
    \hline
  \end{tabular}
\end{table}

\begin{figure}
  \center
  \resizebox{.9\hsize}{!}{\includegraphics[]{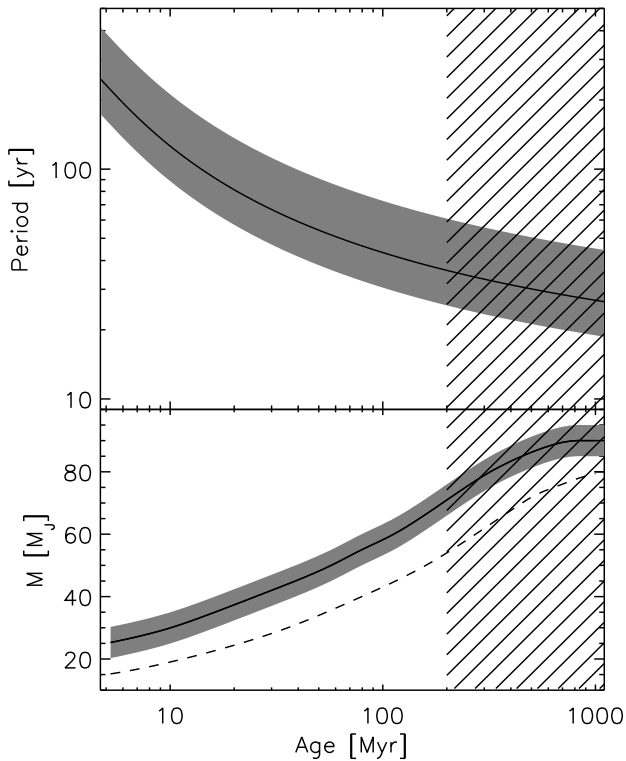}}
  \caption{ \emph{Top panel:} Estimated orbital period as a function
    of age for 2M0041A+B (solid line). The grey region indicates
    1$\sigma$ uncertainties (see text). \emph{Bottom panel:} Estimated
    mass as a function of age. The solid line shows the mass of
    component A, the dashed line is for component B, uncertainties are
    shown for component A only. The hatched area marks the region
    excluded from the presence of Li in the spectrum.
    \label{fig:Period.Age}}
\end{figure}

The orbital period occuring if the binary was old is $P \approx
22$\,yr. This is the lower boundary for the period, the real period
must be longer because an age above a few hundred Myr is excluded by
the detection of Li. If 2M0041 has an age of 10\,Myr, the period on
the order of 126\,yr, i.e., a factor of 5 longer. If the system is as
young as 5\,Myr, the period is about 228\,yr, i.e., another factor of
two longer.  We show the estimated period as a function of age in
Fig.\,\ref{fig:Period.Age}. With the photometric information gathered
about this object, the possible orbital period is a steep function of
age.

\section{Summary}
\label{sect:Summary}

In a NaCo image of the nearby young brown dwarf 2M0041, we have
discovered the binarity of this object. The system consists of an M6.5
primary and a secondary of spectral type M9.0. We find a separation of
143\,mas and derive individual $J$, $H$, and $K_s$ magnitudes for both
components. Flux ratio and colors of both components are consistent,
which means that the chances for the secondary being a background
object are very small.

Distance, age, mass, and orbital period are yet unknown, but we can
present possible solutions as a function of age. The object still has
lithium, which means that it is a brown dwarf younger than a few
hundred Myr.  H$\alpha$ and other optical emission lines indicate that
the object may be accreting so that its age may be as low as
10\,Myr. For this age, the period is predicted to be on the order of
125\,yr at a separation of $\sim9$\,AU and a distance of 50\,pc. Age,
semi-major axis, and distance are steep functions of the orbital
period. So far, no parallax measurement is available so that the
distance to the binary is unknown given that we do not certainly know
its age.

Objects of spectral type M9.0 are much fainter at optical wavelengths
than M6.5 so that the secondary contributes negligible background
continuum to the H$\alpha$ and Li measurement. On the other hand, the
H$\alpha$ emission seen in the combined spectrum may come from either
of the two components, or both, which would affect the measured
accretion rate; if both objects are accreting, the accretion rate per
star would be lower, but the shape and strength of H$\alpha$ still
would indicate accretion. Alternatively, if the accretion timescale is
longer on the secondary, it may be the only accreting object in the
system. In that case, the accretion rate would be higher than the one
originally derived. An image taken at H$\alpha$ or even spatially
resolved spectroscopy can solve this issue.

The discovery of binarity in this young brown dwarf opens the
opportunity to directly determine the mass of two young nearby brown
dwarfs that may be accreting. Without knowing the distance, the age of
the system can be determined from the orbital period. In the next
years, the first estimate of the orbital period will become available.
If it is indeed as young as a few ten Million years, it is the first
object in this mass/age regime for which a direct mass estimate will
become available. With a measured parallax, independent information on
the distance putting strong constraints on the system's parameters
will be given. This system will be an important benchmark for brown
dwarf evolutionary models at young ages.

\begin{acknowledgements}
  We thank an anonymous referee for a very helpful report. Based on
  observations made with the European Southern Observatory, PID
  383.C0708. This publication has made use of the Very-Low-Mass Binaries
  Archive housed at \texttt{http://www.vlmbinaries.org} and maintained by Nick
  Siegler, Chris Gelino, and Adam Burgasser. A.R.  acknowledges research
  funding from the DFG as an Emmy Noether fellow under RE 1664/4-1, AS
  acknowledges financial support from NSF grant AST07-08074.
\end{acknowledgements}

\end{document}